\documentstyle[11pt,newpasp,twoside,epsf]{article}

\begin{document}

\title{Effelsberg HI Survey of Compact High-Velocity Clouds}
\author{Tobias Westmeier, Christian Br\"uns, J\"urgen Kerp}
\affil{Radioastronomisches Institut der Universit\"at Bonn, Auf dem H\"ugel 71, 53121 Bonn, Germany}

\begin{abstract}
We studied 11 compact high-velocity clouds (CHVCs) in the 21-cm line emission of neutral hydrogen with the 100-m 
telescope in Effelsberg. We find that most of our CHVCs are not spherically-symmetric as we would expect in case of 
a non-interacting, intergalactic population. Instead, many CHVCs reveal a complex morphology suggesting that they are 
disturbed by ram-pressure interaction with an ambient medium. Thus, CHVCs are presumably located in the neighborhood 
of the Milky Way instead of being spread across the entire Local Group.
\end{abstract}

\section{Introduction}

High-velocity clouds (HVCs) were discovered by Muller, Oort, \& Raimond (1963) in the 21-cm line emission of 
neutral atomic hydrogen (HI). HVCs are gas clouds which are characterized by high radial velocities incompatible with 
a participation in Galactic rotation. Braun \& Burton (1999) introduced a subclass of isolated HVCs with small 
angular sizes of less than $2^{\circ}$ FWHM. They compiled a catalog of 66 of these so-called compact high-velocity 
clouds (CHVCs) from the Leiden/Dwingeloo Survey of Galactic neutral hydrogen (Hartmann \& Burton 1997) and proposed 
that their statistical properties were consistent with a distribution throughout the entire Local Group with 
distances of the order of $1 \; {\rm Mpc}$. The typical sizes of CHVCs would then be of the order of $15 \; {\rm kpc}$ 
with typical HI masses of a few times $10^7 \; {\rm M_{\odot}}$.

We mapped 11 CHVCs from the Braun \& Burton (1999) and de Heij, Braun, \& Burton (2002) catalogs in 21-cm line 
emission with the 100-m telescope in Effelsberg to validate the results of Braun \& Burton (1999). The high angular 
resolution of $9'$ HPBW allows us to extract the HI structure of these 11 CHVCs in much more detail than the 
previous Leiden/Dwingeloo Survey data. In section~2 we describe the sample selection and data acquisition, 
section~3 outlines our major results, and section~4 includes the summary and conclusions.

\begin{figure}
  \plotone{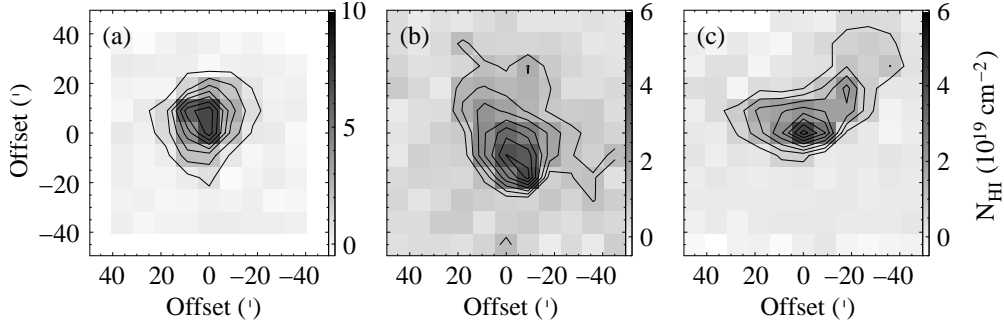}
  \caption{\label{fig_overview}HI Column density maps of (a) spherically-symmetric CHVC 148$-$82, 
  (b) head-tail CHVC 040$+$01 and (c) bow-shock shaped CHVC 172$-$60. Contour levels range from 
  $N_{\rm HI} = 5 \cdot 10^{18} \; {\rm cm^{-2}}$ ($1 \cdot 10^{19} \; {\rm cm^{-2}}$ in (b)) in steps of 
  $5 \cdot 10^{18} \; {\rm cm^{-2}}$.}
\end{figure}

\section{Sample selection and data acquisition}

The 11 studied CHVCs were selected on the basis of earlier, less sensitive Effelsberg observations by two criteria. 
They still had to be classified as CHVCs in the improved catalog by de Heij et~al. (2002), and the HI peak column 
density ratio between the Effelsberg data and the Leiden/Dwingeloo Survey had to be $N_{\rm Eff} / N_{\rm LDS} \ge 3$ 
so that only the most compact CHVCs were selected for reinvestigation.

In general, each object was mapped with $11 \times 11$ spectra on a $9'$ grid (beam-by-beam sampling). 
We reach a sensitivity of $\sigma_{\rm rms} \sim 50 \; {\rm mK}$ at $2.6 \; {\rm km \, s^{-1}}$ velocity resolution. 
These maps allow us to investigate the morphology of our 11 CHVCs and the overall distribution of radial velocities 
and linewidths of the HI gas. In addition, we observed spectra along an appropriate axis of each CHVC with a longer 
integration time and a smaller separation of $4.5'$ or $6.4'$ between the spectra. Along these deep profiles we reach 
a sensitivity of $\sigma_{\rm rms} \sim 30 \; {\rm mK}$, allowing us to obtain the column density profile in more 
detail.

\section{Results}

Table \ref{chvc_table} summarizes the extracted physical parameters of the observed CHVCs, derived by fitting a 
Gaussian to the spectral lines. Radial velocities are mainly negative which is simply a selection effect. 
The mean linewidth of our 11 CHVCs is $\langle \Delta v \rangle = 23 \pm 7 \; {\rm km \, s^{-1}}$, the mean peak 
brightness temperature is $\langle T_{\rm B} \rangle = 1.6 \pm 0.8 \; {\rm K}$, and the average peak column density 
is $\langle N_{\rm HI} \rangle = (5.3 \pm 1.6) \cdot 10^{19} \; {\rm cm^{-2}}$. In some cases, we find evidence for 
two distinct gas phases in the spectral lines, indicating the existence of a core of cold neutral gas embedded in 
a diffuse envelope of warm neutral gas.

Among the 11 investigated objects, only CHVC 148$-$82 appears to be spherically-symmetric (fig. \ref{fig_overview} 
(a)). All other clouds have a more or less complex morphology. 4 CHVCs reveal a head-tail structure which indicates 
that their envelopes of diffuse, warm gas are stripped off by ram-pressure interaction with an ambient medium. The 
conception of an interaction process is also supported by the distribution of radial velocities and linewidths across 
all 4 clouds. Fig. \ref{fig_overview} (b) shows CHVC 040$+$01 as an example for a pronounced head-tail structure.

\begin{figure}
\plotone{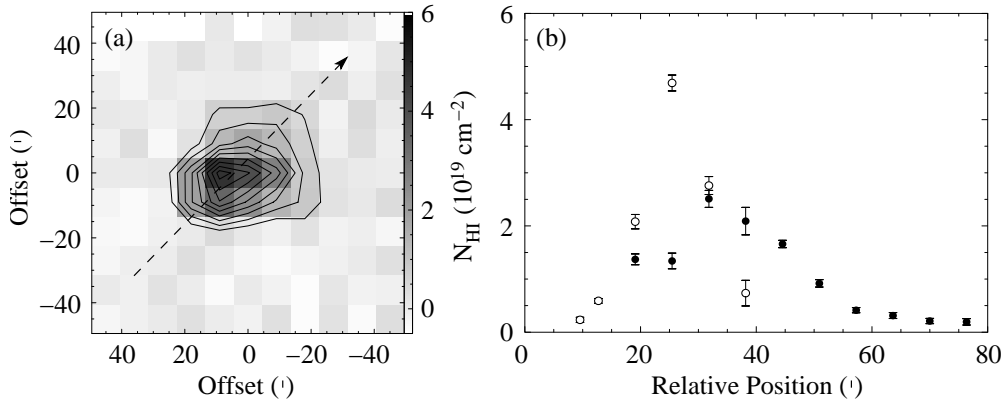}
\caption{\label{fig_chvc01}CHVC 017$-$25. (a) HI column density map. Contours range from 
$N_{\rm HI} = 5 \cdot 10^{18} \; {\rm cm^{-2}}$ in steps of $5 \cdot 10^{18} \; {\rm cm^{-2}}$. The 
dashed arrow marks the position of the deep profile. (b) HI column density distribution of cold 
(open circles) and warm (filled circles) neutral medium along the profile.}
\end{figure}

Another example for a head-tail CHVC is given in fig. \ref{fig_chvc01} (a). CHVC 017$-$25 is a relatively compact 
object, and its slightly asymmetric shape already indicates that gas might have been stripped off the cloud, creating a 
diffuse, faint tail in north-western direction. Furthermore, CHVC 017$-$25 shows a clear two-component structure in the 
HI line profiles. Spectral lines seem to disclose a superposition of a narrow Gaussian component of cold gas 
and a broad component of warm gas. The cold component shows line widths of about $7 \; {\rm km \, s^{-1}}$ FWHM, resulting 
in an upper limit for the gas temperature of roughly $1000 \; {\rm K}$. The warm component discloses much larger line widths 
around $20 \; {\rm km \, s^{-1}}$ FWHM, indicating an upper limit for the gas temperature of about $9000 \; {\rm K}$. Both gas 
components can be investigated separately by a Gaussian decomposition of the spectral lines. Fig. \ref{fig_chvc01} (b) shows 
the distribution of HI column densities of the cold (open circles) and warm (filled circles) gas along the profile indicated 
by the dashed arrow in fig. \ref{fig_chvc01} (a). One can clearly distinguish a compact cold core surrounded by an extended, 
diffuse envelope of warm gas. Furthermore, the cold core and the warm envelope are spatially separated from each other. At 
the south-eastern edge of the cloud, only the cold gas can be traced while the warm gas forms an extended, faint 
tail in north-western direction. These results suggest that the diffuse envelope of warm gas is currently being stripped 
off the compact cold core of CHVC 017$-$25 by ram-pressure interaction with an ambient medium.

Two CHVCs from our sample show up with a bow-shock shaped structure which again indicates the existence of an ambient 
medium. Fig. \ref{fig_overview} (c) shows CHVC 172$-$60 as an example for a bow-shock shaped object. In both CHVCs, radial 
velocities indicate that the gas is decelerated at the presumable front of the cloud. At the same time, linewidths 
increase, indicating higher turbulence or heating of the gas. Again, these observations can best be explained by 
ram-pressure interaction with a medium surrounding the CHVCs.

The remaining 4 CHVCs from our sample reveal an irregular shape. Some of these irregular clouds have large 
radial velocity gradients, suggesting a rotation. In all cases, the irregular morphology again indicates a disturbance of 
the objects.

\begin{table}
\caption{\label{chvc_table}Physical parameters of the 11 observed CHVCs. $l$ and $b$ are the Galactic longitude and 
latitude of the column density maximum, $v_{\rm LSR}$ and $v_{\rm GSR}$ the column density weighted average radial 
velocities in LSR and GSR frames, $\Delta v$ the average linewidth (FWHM), $T_{\rm B}$ the observed peak brightness 
temperature, and $N_{\rm HI}$ the HI peak column density.}
\begin{center}
\begin{tabular}{lccccc}
\tableline
Name & $v_{\rm LSR}$ & $v_{\rm GSR}$ & $\Delta v$ & $T_{\rm B}$ & $N_{\rm HI}$ \\
(CHVC $l \pm b$) & ($\rm km \, s^{-1}$) & ($\rm km \, s^{-1}$) & ($\rm km \, s^{-1}$) & (K) & ($10^{19} \; {\rm cm^{-2}}$) \\
\tableline
CHVC 016.8$-$25.2    &    $-228$ &   $-171$ &      14 &       3.1 &       5.6 \\
CHVC 032.1$-$30.7    &    $-308$ &   $-207$ &      30 &       1.3 &       6.0 \\
CHVC 039.0$-$33.2    &    $-262$ &   $-147$ &      22 &       1.6 &       8.0 \\
CHVC 039.9$+$00.6    &    $-278$ &   $-137$ &      32 &       0.7 &       4.5 \\
CHVC 050.4$-$68.4    &    $-195$ &   $-133$ &      27 &       1.3 &       4.7 \\
CHVC 147.5$-$82.3    &    $-269$ &   $-254$ &      22 &       2.2 &       8.0 \\
CHVC 157.1$+$02.9    &    $-184$ &   $ -98$ &      22 &       0.9 &       3.6 \\
CHVC 172.1$-$59.6    &    $-235$ &   $-219$ &      28 &       0.9 &       4.2 \\
CHVC 218.1$+$29.0    &    $+145$ &   $ +27$ &       6 &       2.8 &       3.2 \\
CHVC 220.5$-$88.2    &    $-258$ &   $-263$ &      22 &       1.0 &       3.7 \\
CHVC 357.8$+$12.4    &    $-159$ &   $-167$ &      27 &       1.5 &       6.4 \\
\tableline
\tableline
\end{tabular}
\end{center}
\end{table}

\section{Summary and conclusions}

We showed that 10 of our 11 investigated CHVCs reveal a complex structure and that many of them show signs for 
ram-pressure interaction with a surrounding medium. This is in opposite to the original idea of CHVCs being the gaseous 
counterparts of primordial Dark-Matter halos spread across the entire Local Group. Instead, the observed ram-pressure 
effects and different distance estimates (Westmeier 2003) indicate that CHVCs constitute a circumgalactic population with 
typical distances of the order of $100 \; {\rm kpc}$. At such distances, the observed interactions could be caused by 
an extended Galactic halo gas. Typical HI masses of CHVCs would then be of the order of a few times $10^5 \; {\rm M_{\odot}}$ 
with typical sizes of about $1 \; {\rm kpc}$.

\acknowledgements

This work is based on observations with the 100-m telescope of the MPIfR (Max-Planck-Institut f\"ur Radioastronomie) 
at Effelsberg.


\begin{references}
\reference Braun, R. \& Burton, W. B. 1999, \aap, 341, 437
\reference Hartmann, D. \& Burton, W. B. 1997, Atlas of Galactic Neutral Hydrogen, Cambridge University Press
\reference de Heij, V., Braun, R., \& Burton, W. B. 2002, \aap, 391, 159
\reference Muller, C. A., Oort, J. H., \& Raimond, E. 1963, C.R.Acad.Sci.Paris, 257, 1661
\reference Westmeier, T. 2003, Diploma Thesis, Universit\"at Bonn
\end{references}
\end{document}